 \documentclass[11pt]{article}
\usepackage{graphicx} % standard LaTeX graphics tool
\usepackage{epsfig}
\usepackage{amssymb}
\usepackage{epstopdf}
\usepackage{amsmath}
\newtheorem{theorem}{Theorem}
\newtheorem{definition}{Definition}

\title{\begin{center}Stereotype bias:   a simple formal model \end{center}}
\author{Fran\c{c}ois Bavaud\\  \\
  Department of Computer Science and Mathematical Methods\\
Department of Geography\\
University of Lausanne,  Switzerland}

\date{}

\begin{document}

\maketitle             

\begin{abstract}
Minimizing the relative inertia of a statistical group with respect to the inertia of the overall sample defines  an unique point, the in-focus, which constitutes a  context-dependent measure of typical group tendency, biased in comparison to the group centroid. Maximizing the relative inertia yields an unique out-focal point, polarized in the reverse direction. This mechanism   evokes the relative variability reduction of the outgroup reported in Social Psychology, and the stereotypic-like behavior of the in-focus, whose bias vanishes if the outgroup is constituted of a single individual. In this picture, the out-focus plays the role of an anti-stereotypical position, identical to the in-focus of the complementary group.  \end{abstract}


\section{Introduction}
The expression {\em  ``typical features of a group"} is ambiguous: it might either refer to a multivariate indicator of group central tendency, that is to an unbiased  {\em group centroid}, or to the  distinctive, unique Ê characteristic tendencies of the group,  contrasting  the features observed in the other groups or in the complete sample, in which case it constitutes a caricatural  {\em stereotype}. 

A  stereotype summarizes the features of a whole group into a single profile (variability reduction). Its profile is generally distinct from the group centroid, pushed away from 
the overall centroid of the whole population or context under consideration  (bias or polarization). 

These two characteristics of stereotypy 
have been largely  reported in Social Psychology, in particular  with reference to the outgroup: 
\begin{enumerate}
  \item[{\bf 1)}] people tend to minimize the differences between
outgroup members, while being inclined to perceive their own
group as  made of an heterogeneous set of unique individuals  ({\em outgroup relative homogeneity}): ``They all look alike but we don't". See e.g. Quattrone and Jones  (1980),  Taylor et al. (1978), Park and Hastie  (1987), Mullen and  Hu (1989), and references therein. 
  \item[{\bf 2)}] people tend to exaggerate the typical traits of the outgoup, and to enhance the contrasts between members of different groups
({\em stereotype polarization or bias effect}). See e.g. Turner (1975),  Hopkins and Cable (2001), Hogg et al. (2004),  Realo et al. (2009),  and references therein. 
\end{enumerate}

This paper  presents a simple, formal, principled mechanism linking the two aforementioned aspects of stereotypy. Specifically, we show that minimizing the  {\em relative  group dispersion} defines an unique point in the feature space, called the {\em in-focus} (Theorem \ref{iafp}), manifesting  the 
exaggeration or polarization effect  expected from a stereotype  (Theorem \ref{diafpolar}). The polarization can be qualified as {\em fair}, in the sense it vanishes for a group formed of a single individual, which thus coincides with its own stereotype. Increasing the group dispersion increases the polarization effect, as shown by (\ref{adepsilon}) and (\ref{trucad}).

Furthermore, maximizing the relative  group dispersion yields another unique point conjugate to the in-focus, the {\em out-focus},  bearing the characteristics of an anti-stereotype or {\em antitype}. 
The group out-focus coincides with the in-focus of the {\em complementary group} (Theorem \ref{theo3}), as illustrated on the U.S. Congressmen data (Section \ref{exuscong}).

Section  \ref{fucon} attempts to justify the position of the in-focus in  a decision-theoretical setup, and underlines the connections with the   {\em meta-contrast} model Êand  simulations of Salzarulo (2006), whose work initially triggered the present research. 
 
\section{The formal model}

\subsection{Definitions and notations: Huygens principle}
Consider a totality $n$ individuals, denoted $i=1,\ldots,n$, characterized by a   multivariate   profile
of $p$ features  $x_{ik}$ with $k=1,\ldots,p$. These features define a squared Euclidean distance between individuals  
\begin{equation}
\label{sqeuc}
D_{ij}=\sum_{k=1}^p (x_{ik}-x_{jk})^2\qquad\qquad i,j=1,\ldots, n\enspace .
\end{equation}
For generality sake, we assume that individuals possess {\em weights} $f_i>0$, with $\sum_{i=1}^nf_i=1$. The uniform weighting obtains as $f_i=1/n$. Also, consider a profile $a\in \mathbb{R}^p$, which might or might not correspond to the features of  an existing individual.  Huygens principle  consist of the identities
\begin{equation}
\label{huy}
\Delta_f^a=\sum_{i=1}^n f_i D_{ia}=\Delta_f +D_{fa} \qquad\qquad\qquad \Delta_f=\frac12\sum_{ij}f_if_jD_{ij}
\end{equation}
where $\Delta_f^a$ is the inertia relative to the reference point $a$, $ D_{fa} $ is the squared distance between the point $a$ and the centroid $\bar{x}_f=\sum_i f_i x_i$, and $\Delta_f=\Delta_f^{\bar{x}_f}$ is the inertia relative to the centroid. In particular, (\ref{huy}) shows that $\Delta_f^a$ attains its minimum $\Delta_f$ for  $a=\bar{x}_f$. 

Now consider a group $g$ of individuals. A group is specified by the individuals it contains, that is, in full generality, by a distribution $g_i\ge 0$ with $\sum_{i=1}^n g_i=1$. In most situations, the support of $g$ (that is the set of individuals for which $g_i>0$)  is a strict subset of the complete set of the $n$ individuals, but this restriction is not necessary. We however assume that the group  centroid $\bar{x}_g=\sum_i g_i x_i$ differs from the overall  centroid $\bar{x}_f$, that is $D_{fg}>0$. As before, 
$\Delta_g^a$ takes on its minimum value $\Delta_g$ for $a=\bar{x}_g$. 

\subsection{The relative dispersion}
\label{secredi}
\begin{figure}
\begin{center}
\includegraphics[width=3in]{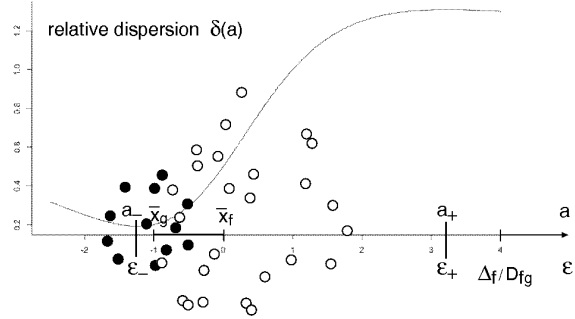}
\caption{Abscissa: absolute coordinates $a$ and relative coordinates
$\epsilon$, along the line (\ref{adepsilon}) passing through the group centroid $\bar{x}_g$ (associated to the black objects) and the overall centroid $\bar{x}_f$ (associated to the black and white objects). Ordinate:  the relative dispersion $\delta(a)$ is minimum for the  in-focus points $a_-$, maximum 
for   the  out-focus points $a_+$,  and 
tends to unity as $\epsilon\to\pm\infty$.}
\label{fig12}
\end{center}
\end{figure}

\begin{definition}[Relative dispersion]
\label{reldisp}
 The {\em relative dispersion} of group $g$ in context $f$, relatively to the reference  point $a$ is  
\begin{equation}
\label{delredis}
\delta(a)=\delta(a|g,f)=\frac{\Delta_g^a}{\Delta_f^a}=\frac{\Delta_g+D_{ga}}{\Delta_f+D_{f a}}\enspace .
\end{equation}
\end{definition}
The relative dispersion measures the  disparity  or  heterogeneity  in group $g$, in units determined by the   overall heterogeneity, {\em as assessed from some reference point of view $a$}.  Varying the point of view enables to tune, within some limits, the apparent, perceived relative
 heterogeneity of the group $g$. Remarkably enough, the relative dispersion $\delta(a)$ is finite everywhere, and possesses an unique minimum $a_-$ as well as an unique maximum $a_+$:

 \begin{theorem}[In- and out-focus points]
 \label{iafp}
 Both the {\bf in-focus point} $\mathbf{a_-}$ minimizing $\delta(a)$,  and the  {\bf out-focus point} $\mathbf{a_+}$  maximizing $\delta(a)$ are  unique, and given by $a_\pm:=a(\epsilon_\pm)$, with
\begin{equation}
\label{adepsilon}
a(\epsilon):=\bar{x}_f+\epsilon(\bar{x}_f-\bar{x}_g)\qquad\qquad
\epsilon_\pm =  \frac{1}{2D_{f g}}(b_{f g}\pm\sqrt{b_{f g}^2+4 \Delta_f D_{f g}})
\end{equation}
 where $b_{f g}:=\Delta_f-\Delta_g-D_{f g}$. 
   \end{theorem}
By theorem \ref{iafp}, the in- and out-focus points lie on the line joining centroids $\bar{x}_f$ and $\bar{x}_g$. In-focus  polarization    occurs if $a_-$ lies ``on $\bar{x}_g$ side", that is if 
$\epsilon_-\le -1$, as in Figure \ref{fig12}. Similarly, out-focus  polarization occurs if $a_+$ lies ``on $\bar{x}_f$ side", that is if 
$\epsilon_+\ge 0$. Theorem \ref{diafpolar} insures this to be always the case. 
   
\begin{theorem}[Polarization]
\label{diafpolar}
In- and out-focus points $a_\pm$ fall outside the interval $[\bar{x}_g, \bar{x}_f]$, as in  figure  \ref{fig12}. Specifically,
  \begin{eqnarray}
\label{ass2}
\epsilon_-\le -1
\qquad\qquad\qquad
0<\epsilon_+\le \frac{\Delta_f}{D_{f g}}
\end{eqnarray}
where both inequalities are attained iff $\Delta_g=0$, as in the case of a singleton, dispersion-free group. For small  $\Delta_g$, 
  \begin{eqnarray}
  \label{trucad}
\epsilon_-& = &-1-[\frac{1}{\Delta_f+D_{f g}}]\: \Delta_g+0(\Delta_g^2)\\
\epsilon_+& = &\frac{\Delta_f}{D_{f g}}-[\frac{\Delta_f}{D_{f g}(\Delta_f+D_{f g})}]\: \Delta_g+0(\Delta_g^2) 
\end{eqnarray}
\end{theorem}

\noindent  {\bf Proofs}: let  $e:=\|\bar{x}_f-a\|$ be fixed, and consider the angle $\alpha$ between $\bar{x}_g$ and $a$ as measured from $\bar{x}_f$.  By the  cosine theorem,
  \begin{eqnarray*}
\frac{\Delta_g+D_{ga}}{\Delta_f+D_{f a}}=\frac{\Delta_g+D_{f g}+e^2-2\sqrt{D_{fg}}\:  e \cos\alpha}{\Delta_f+e^2}
 \end{eqnarray*}
which is maximum for $\alpha=180^\circ$ and  minimum for $\alpha=0^\circ$. In both cases, the extremum $a$ lies on the line passing through $\bar{x}_g$ and $ \bar{x}_f$, i.e. is of the form $a(\epsilon)$ in 
(\ref{adepsilon}), with relative dispersion $(\Delta_g+D_{f g}(\epsilon+1)^2 )/(\Delta_f+ D_{f g}\epsilon^2)$. Setting to zero its derivative in $\epsilon$ yields $D_{f g}\epsilon^2-b_{f g}\epsilon -\Delta_f=0$, with solutions $\epsilon_\pm$ (with the correct sign)  given by  (\ref{adepsilon}).  Furthermore, it is easy to show that, for $D_{f g}$ and $\Delta_f$ fixed, both expressions $\epsilon_-$ and $\epsilon+$ are decreasing in $\Delta_g$, and take on their maximum value (\ref{ass2}) for
 $\Delta_g=0$.    $\Box$
  
\subsection{Other expressions}
The following features-based expression may be computationally useful:
  \begin{eqnarray}
a_{\pm}=\sum_i\alpha_i(\epsilon_\pm)\:  x_i
\qquad\qquad \alpha_i(\epsilon):=(1+\epsilon)f_i -\epsilon g_i\: .
 \end{eqnarray}
$\alpha(\epsilon)$ is a {\em signed} distribution, that is normalized to unity but not necessarily non-negative.

Also, twice application of Huygens decomposition (or direct manipulation of  the features) demonstrates the distance-based identities
\begin{eqnarray*}
D_{f g}=-\frac12\sum_{ij}(f_i-g_i)(f_j-g_j)D_{ij}
\qquad \qquad b_{f g}=\sum_{ij}f_i(f_j-g_j)D_{ij}\: .
 \end{eqnarray*}
 Finally, define the {\em squared polarization ratio}  as
\begin{displaymath}
\frac{D_{{a_-}{a_+}}}{D_{fg}} = (\epsilon_+-\epsilon_-)^2=
(1+\frac{\Delta_f+\Delta_g}{D_{fg}})^2-
4\frac{\Delta_f \Delta_g}{D_{fg}^2}\: \ge\: 1
\end{displaymath}
which shows the polarization   to  increase with each of the inertias $\Delta_f$ and $\Delta_g$. In particular,  the 
polarization ratio takes on its minimum value unity iff $\Delta_g=\Delta_f=0$, and $a_{-}=\bar{x}_g$ iff 
$\Delta_g=0$, as shown by (\ref{ass2}).

\subsection{Illustration: U.S. Congressmen}
\label{exuscong}
In  the legislature 1984, a number of $p=16$  ``key votes" from $n=435$ US Congressmen, comprising
$n_R=168$ Republicans and $n_D=267$ Democrats,  have been coded as 1 (``yea") or 0 
(``nay")\footnote{http://archive.ics.uci.edu/ml/machine-learning-databases/voting-records/ }.  Missing values have been replaced by the average value inside the affiliated political group.

\begin{figure}[h]
\begin{center}
\includegraphics[width=3.5in]{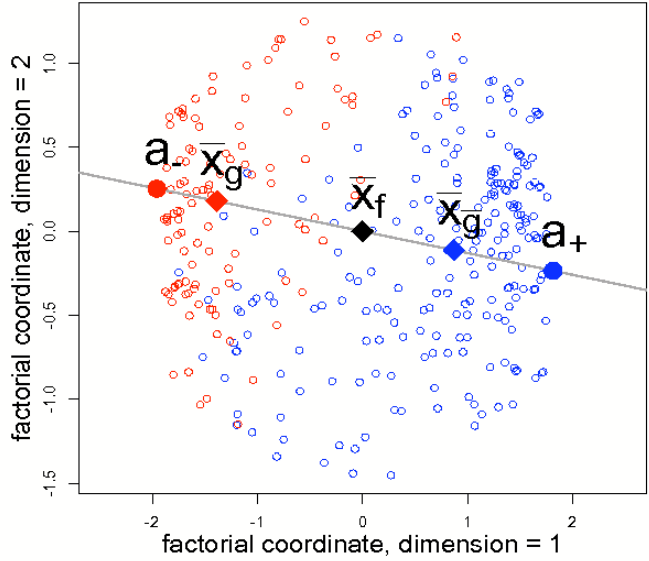}
\caption{First MDS coordinates obtained from the squared distances between U.S. Congressmen (legislature 1984), expressing $58\%$ of the overall dispersion $\Delta_f$. Red  circles depict the positions of the  Republicans, and blue circles those of the Democrats. $\bar{x}_f$ is the overall centroid, $\bar{x}_g$ is the Republicans centroid,
$\bar{x}_{\bar{g}}$ is  the Democrats centroid.  $a_-$ is the in-focus relative to the Republicans, and $a_{+}$ the corresponding out-focus; see also Section \ref{fiar}.}
\label{rep_demo}
\end{center}
\end{figure}

The overall, Republican and Democrat  distributions read  respectively  as 
\begin{equation}
\label{congrecen}
f_i=\frac{1}{n}\qquad\qquad  g_i=\frac{I(i\in R)}{n_R}
\qquad\qquad
\bar{g}_i=\frac{I(i\in D)}{n_D}\enspace .
\end{equation}
where $I(A)$ denotes the characteristic function of event $A$. After computation of the  squared Euclidean distances (\ref{sqeuc})  from the  Congressmen votes,  a classical multidimensional scaling (MDS) has been performed with uniform weighting of the individuals (see e.g. Mardia et al. 1979) to obtain the  factorial  coordinates expressing a maximum amount of  the overall  dispersion $\Delta_f$ (Figure \ref{rep_demo}).  With $D_{fg}=1.98$, $\Delta_f=3.67$ and $\Delta_g=1.89$,  the polarization ratio is $|\epsilon_+-\epsilon_-|= 2.72$.

\subsection{Complementary group}
\label{fiar}
By construction (Section \ref{secredi}),  the  in-focus   $a_-$ is the point of view under which group $g$ appears, relatively to the {\em ground} or {\em context} formed by the complete set of individuals $f$,  {\em as  homogeneous as possible}, and  turns out to constitute a credible  candidate for representing a stereotypical  value. By contrast, the out-focus $a_+$ is the point of view maximally respectful of the features diversity in $g$, and behaves as an  {\em antitype} - in the sense of ``anti-stereotypical". 

In the example of  Figure \ref{rep_demo}, the out-focus $a_+$ of the Republicans seems to be equally qualified to  represent the 
 in-focus of the Democrats (not drawn on the Figure). As a matter of fact, the two points {\em coincide}, as  justified by Definition \ref{def2} and Theorem \ref{theo3} below.

\begin{definition}
\label{def2}
The group  {\em complementary} to  group $g$ in context $f$ is defined by a  normed distribution  $\bar{g}$,  which, mixed with $g$, reproduces $f$, in the sense $\rho g+(1-\rho)\bar{g}=f$, or equivalently 
\begin{equation}
\label{compgroup}
\bar{g}_i=\frac{f_i-\rho g_i}{1-\rho}\qquad\quad\mbox{for some $\rho\in(0,\rho_{\max}]$}
\quad \mbox{with}Ê\quad\rho_{\max}:=\min_i (f_i/g_i)\enspace .
\end{equation}
\end{definition}
The definition of $\rho_{\max}$  insures the non-negativity of $\bar{g}$. For instance, the Democrats group $\bar{g}$ in (\ref{congrecen}) is complementary to the Republicans group $g$ since 
$\rho g+(1-\rho)\bar{g}=f$ with $\rho=n_R/n=.39$, which turns out to be equal to its maximum value $\rho_{\max}$.

\begin{theorem}
\label{theo3}
The  out-focus point   $a_+^g$ for group $g$ is the in-focus point $ a_-^{\bar{g}}$ of any group 
 $\bar{g}$ complementary to $g$.
\end{theorem}

{\bf Proof:}  substituting  $f=\rho g+(1-\rho)\bar{g}$ in (\ref{huy}) and developing  the first identity 
demonstrates $\Delta_f^a=\rho(\Delta_g+D_{ga})+(1-\rho)(\Delta_{\bar{g}}+D_{\bar{g}a})$,   that is 
$\rho\:  \delta(a|g,f)+(1-\rho)\: \delta(a|\bar{g},f)=1$. Hence, $\rho$, $f$ and $g$  being fixed,   maximizing  $\delta(a|g,f)$ amounts in minimizing  $\delta(a|{\bar{g}},f)$. $\Box$

\vspace{0.1cm}

Note the identity 
\begin{equation}
\label{equivegfgbar}
\delta'(a)=\delta(a|g,\bar{g})= \frac{\Delta_g^a}{\Delta_{\bar{g}}^a}=\frac{1-\rho}{\frac{\Delta_f^a}{\Delta_g^a}-\rho}
\end{equation}
which  demonstrates that  extremalizing the relative dispersion $\delta(a)=\delta(a|g,f)$ or its variant $\delta'(a)$ yields  
  the {\em same} solutions.

\section{Further connections}
\label{fucon}

\subsection{Decision theory}
The following argument constitutes a first attempt towards a derivation  of the in-focus in a decision-theoretical framework. 
Consider the decision rule  {\em ``attribute individual $i$ either to group $g$ with probability $P(g|i)=\exp(-\beta D_{ia})$ or to the overall set $f$ with probability $P(f|i)=1-P(g|i)$"}, where $\beta>0$ is a parameter controlling the decay of the exponential   and $a$ a point to be chosen wisely. The probability of misidentifying an individual of $g$ (miss) is $P(f|g)=\sum_i g_i P(f|i)$, and the probability of correctly 
identifying an individual of $f$ as such (correct rejection) is $P(f|f)=\sum_i f_i P(f|i)$. In the limit of large spread, the ratio of these quantities becomes
 \begin{displaymath}
\lim_{\beta\to 0}\frac{P(\mbox{miss})}{P(\mbox{co.re.})}
=\lim_{\beta\to 0}\frac{P(f|g)}{P(f|f)} 
=\lim_{\beta\to 0} \frac{\beta \sum_i g_i D_{ia}+0(\beta^2)}
{\beta \sum_i f_i D_{ia}+0(\beta^2)}=\frac{\Delta_g^a}{\Delta_f^a}=\delta(a)\enspace .
\end{displaymath} 
In this context, the  probability of miss is minimized by the group centroid $\bar{x}_g$, while the
ratio of the probabilities  ``miss over correct rejection" is minimized by the in-focus $a_{-}$ - a  somewhat  intriguing result to be further investigated. 

\subsection{Subtractive combinations}
Instead of studying the relative dispersion {\em ratio} (\ref{delredis}), one can consider the {\em subtractive  combination}  of the  form  
\begin{equation}
\label{subsim}
\gamma(a)=A\: \Delta_g^a-B\: \Delta_f^a \qquad\qquad A,B\in \mathbb{R}
\enspace .
\end{equation}
If $A+B\neq0$, the two parameters can be normalized as $A=1-\lambda$ and $B=\lambda$. 
For $ \lambda\neq0.5$, $\gamma(a)$ possesses a unique  bounded extremum at $\epsilon=(\lambda-1)/(1-2\lambda)$ (following parameterization (\ref{adepsilon})), which turns out to be a minimum for 
$\lambda<0.5$ and a maximum for $\lambda>0.5$; no bounded extremum exists for $ \lambda=0.5$. 
If $A+B=0$,   $\gamma(a)$ possesses a  unique  bounded extremum at the mid-point $\epsilon=-0.5$,  which constitutes a minimum for $A>0$ and a maximum for $A<0$. In any case, the position of the extremum does {\em not} depend upon the dispersions $\Delta_f$ and $\Delta_g$. Similar results are obtained when replacing $f$ by $\bar{g}$ in (\ref{subsim}).

\subsection{Meta-contrast ratio and prototypicality function}
More interesting, and considerably more involved is the study of the following 
function, appearing in the  framework of the {\em self-categorization theory} (where groups are not given a priori), proposed by Salzarulo (2006), and referred to him (up to a sign)  as the {\em prototypicality function}: 
\begin{equation}
\label{subinvo}
\Gamma(a)=(1-\lambda)\: \Delta_{g(a)}^a-\lambda\: \Delta_{\bar{g}(a)}^a  \qquad\qquad\lambda\in[0,1]
\end{equation}
Here $g_i(a)= \exp(-\beta D_{ia})/Z(a)$, where $Z(a)$ is the normalization constant: in this approach,   the very composition of group $g$  depends on the distance of its constituents to  $a$.
 Also, $\bar{g}(a)$ is of the form (\ref{compgroup}) with  $\rho(a)= Z(a)/nÊ<1$.

The function (\ref{subinvo}) is primarily meant as an improved variant of the  {\em metacontrast ratio} (Haslam and Turner 1995; Turner et al. 1987; Oakes et al. 1994),  measuring the relative differences between individuals, and aimed at predicting to which extent a given individual will be perceived as belonging to the  subject group. 

The  highly non-linear properties of of $\Gamma(a)$, whose minima are interpreted as prototypical positions, 
can be built on to run
dynamical numerical simulations exhibiting groups formation and destruction, in the context of opinion formation, for various values of $\lambda$ and $\beta$. In particular, two agents initially categorizing themselves as different can perceive themselves as belonging to the same group in presence of a third agent distant from them; also, fitting experimental data is possible, as those of Haslam and Turner (1995), satisfactorily reproduced with  $\lambda=.08$ and $\beta=7.7$, on a one-dimensional opinion space $x\in [0,1]$.  See Salzarulo (2006) for more details.

 \section{Discussion and conclusion}
The mechanism relating the minimization of the relative dispersion to the polarization of the in-focus is entirely mathematical, and relevant to the construction of statistically biased, context-dependent measures of central (or ``typical") tendency. However, the parallel with a few predominant themes of Social Psychology seems striking, and we did not resist the temptation to interpret the in-focus as a stereotype, and the statistical group $g$ as an outgroup. The extent to which 
 the metaphor is legitimate  is be judged within  Social Psychology. Among the points potentially stimulating, let us mention the question of the {\em identification of  the ingroup}, of which both the context $f$ and the complementary $\bar{g}$ are  legitimate candidates - with similar if not identical effects, in view of Theorem  \ref{theo3} and (\ref{equivegfgbar}). 
 
 The formalism we have used is both general, that is using weighted groups allowing fuzzy memberships, and classical, that is using squared Euclidean distances as measures of dissimilarities. Euclidean distances  permit, in contrast to other dissimilarities, to extract the original features through MDS (up to a rotation in the features space); they furthermore additively decompose accordingly to  Huygens principles, the use of which has been crucial  in the present  paper.
 
Non-Euclidean dissimilarities, to which alternatives measures of central tendency  are associated,  such as the trimmed mean or the median (e.g. Hampel et al. 1986),  are  perfectly legitimate, and possibly better justified form robustness considerations. The resuting  polarization of the in-focus 
would certainly deserve proper studies, which are however bound to be technically more involved.

\end{document}